\documentclass{revtex4}
\usepackage{amscd,amsmath,amssymb}
\usepackage{bbm,latexsym}
\usepackage[colorlinks=true]{hyperref}
\usepackage{color}

\newcommand{\R}{{\mathbb R}}
\newcommand{\Z}{{\mathbb Z}}
\newcommand{\unity}{{\mathbbm{1}}}

\newcommand{\sfrac}[2]{{\textstyle\frac{#1}{#2}}}
\newcommand{\half}{{\sfrac12}}

\newcommand{\im}{{\mathrm{i}}}

\renewcommand{\and}{{\quad{\rm and}\quad}}
\newcommand{\und}{{\qquad{\rm and}\qquad}}
\renewcommand{\=}{\ =\ }

\newcommand{\ba}{\begin{array}}
\newcommand{\ea}{\end{array}}
\newcommand{\be}{\begin{equation}}
\newcommand{\ee}{\end{equation}}
\newcommand{\bea}{\begin{eqnarray}}
\newcommand{\eea}{\end{eqnarray}}
\newcommand{\bi}{\begin{itemize}}
\newcommand{\ei}{\end{itemize}}

\begin{document}

\begin{flushright}
ITP-UH-17/14
\end{flushright}
\vspace{-4mm}

\title{\LARGE{
Superintegrability of (generalized) Calogero models \\[2pt]
with oscillator or Coulomb potential}
\\[-10pt] \phantom{.}
}
\author{\Large{Tigran Hakobyan}}
\email{tigran.hakobyan@ysu.am}
\affiliation{Yerevan State University, 1 Alex Manoogian St., Yerevan, 0025, Armenia}
\author{\Large{Olaf Lechtenfeld}}
\email{lechtenf@itp.uni-hannover.de}
\affiliation{Leibniz Universit\"at Hannover, Appelstra\ss{}e 2, 30167 Hannover, Germany }
\author{\Large{Armen Nersessian}}
\email{arnerses@ysu.am}
\affiliation{Yerevan State University, 1 Alex Manoogian St., Yerevan, 0025, Armenia}
\affiliation{Tomsk Polytechnic University, Lenin Ave. 30, 634050 Tomsk, Russia}

\begin{abstract} \noindent $\phantom{.}$\\
We deform $N$-dimensional (Euclidean, spherical and hyperbolic) oscillator and Coulomb systems,
replacing their angular degrees of freedom by those of a generalized rational Calogero model.
Using the action-angle description, it is established that maximal superintegrability is retained.
For the rational Calogero model with Coulomb potential, we present all constants of motion via
matrix model reduction. In particular, we construct the analog of the Runge-Lenz vector.
\end{abstract}

\maketitle

\noindent
{\bf 1. Introduction.}
The rational Calogero model \cite{calogero0} and its various generalizations,
based on arbitrary Coxeter root systems \cite{calogero-root} or with
trigonometric and elliptic potentials \cite{cal-tri}),  continue to attract interest.
The reason is their rich integrability structure and their widespread applications
(for a review, see \cite{calogero-review}).
One of the striking features of rational $N$-particle ($A_{N-1}$) Calogero models
is their maximal superintegrability
(i.e. existence of $2N{-}1$ functionally independent constants of motion),
first established by Wojcechowski~\cite{woj83}.
This property is retained for other root systems, and it admits the addition
of an external oscillator potential~\cite{gonera99-2},
which we call the \emph{Calogero-oscillator} system.
Superintegrability has been established also for the hyperbolic Calogero model~\cite{gonera98-1}
and for the relativistic version~\cite{feher10,feher12}
known as the rational Ruijsenaars-Schneider model~\cite{rui}.

One may ask whether other potentials (different from the oscillator one)
may be added to the rational Calogero model without destroying its integrability.
To answer this question, it is fruitful to reinterpret the model as describing
a single nonrelativistic particle moving in $\R^N$ in the presence of a particular potential
given by the root system (e.g.~$A_{N-1}$).
The conformal invariance of this system then suggests passing to spherical coordinates
$(r,\theta_i)$ with $i=1,\dots,N{-1}$ and trying to add a `radial' potential~$V(r)$.
As an important example, Khare~\cite{khare} (see also~\cite{khare99}) proposed
an exactly solvable rational Calogero model with a Coulomb-like radial potential,
to which we will refer as the \emph{Calogero-Coulomb} model.
This was followed by a remark of Calogero~\cite{comment} that \emph{any} radial deformation
of the Calogero model by a rotationally invariant exactly solvable potential remains exactly solvable.

In this Note we establish \emph{maximal superintegrability} for the Calogero-Coulomb and
Calogero-oscillator models associated to any Coxeter root system, as well as for their generalizations
from $\R^N$ to the $N$-sphere~$S^N$ and the (two-sheeted) hyperboloid~$H^N$.
Furthermore, we suggest that these Calogero-type deformations of the $N$-dimensional
oscillator and Coulomb systems are their only modifications preserving maximal superintegrability.

The construction of these models proceeds in two steps.
Firstly, we rewrite the $N$-dimensional oscillator or Coulomb system in spherical coordinates,
and secondly, we replace the angular part of either system (given by the quadratic $SO(N)$ Casimir)
with the angular part of a rational Calogero model (given by its $SL(2,\R)$ Casimir),
which has recently been considered in a series of papers~\cite{sphCal,sph-mat,hlnsy,lny,flp}.
A proof of maximal superintegrability directly follows from the formulation of these systems
in terms of action-angle variables~\cite{hlnsy}.
The latter requires the knowledge of an action for the angular part of the rational Calogero model,
which we get by taking the classical limit of its energy spectrum, as constructed in~\cite{flp}.

As an example, we consider the rational $A_{N-1}$ Calogero model with a Coulomb potential,
for which we present all classical constants of motion via the well known matrix-model reduction 
procedure~\cite{calogero-root,kazhdan}. 
In particular, we determine the analog of the Runge-Lenz vector. 
The quantum integrals are briefly discussed using the Dunkl operator approach~\cite{brink-poly}.
\\

\noindent
{\bf 2. Action-angle variables and superintegrability.}
The description of an integrable system in terms of its action-angle variables
provides a convenient way to determine the hidden symmetries.
Concretely, consider an $N$-dimensional system with action-angle variables
$\bigl(I_i\in\R,\Phi_j\in[0,2\pi)\bigr)$
for $i,j=1,\dots,N$ and a Hamiltonian of the form
\be
{\cal H}\={\cal H}(nI_1{+}mI_2,I_3,\ldots, I_N)
\qquad\text{with}\qquad \{I_i,\Phi_j\}=\delta_{ij}\ ,
\ee
where $n$ and $m$ are integers.
This system features an hidden symmetry, given by the additional constant of motion
\be
I_{\rm hidden}\=A\cos\left(m\Phi_1-n\Phi_2+\alpha\right)\ ,
\ee
with $A(I)$ and $\alpha(I)$ being arbitrary functions of the action variables.
By iteration, the Hamiltonian of a \emph{maximally superintegrable} system should
depend only on one integer linear combination of the action variables,
\be
{\cal H}\={\cal H}(n_1I_1+m_2I_2+\ldots n_N I_N)\ ,
\ee
where $n_1,\ldots, n_N$ are integers. In that case the $\half N(N{-}1)$ functions
\be
I_{ij}\=A_{ij}(I)\cos(n_j\Phi_i-n_i\Phi_j+\alpha_{ij}(I))
\ee
all are constants of motion functionally independent from the Liouville integrals $I_1,\ldots, I_N$.
>From this set we can choose $N{}-1$ functionally independent ones, say $I_{i,i+1}$ with $i=1,\ldots N{-}1$.
Hence, such a system possess $2N{-}1$ functionally independent constants of motion,
i.e. it is maximally superintegrable.

In a recent paper \cite{hlnsy} we have constructed integrable deformations of $N$-dimensional oscillator
and Coulomb systems on Euclidean spaces, spheres and (two-sheeted) hyperboloids, by replacing their
``angular part'' (on $S^{N-1}$) by an arbitrary $(N{-}1)$-dimensional integrable system,
formulated in terms of action-angle variables.
On the Euclidean space $\R^N$, these deformations are defined by
\be
{\cal H}\=\frac{p^{2}_r}{2}+\frac{{\cal I}(I_i)}{r^2}+V(r)
\qquad\text{with}\qquad \{p_r, r\}=1\ ,
\label{2}
\ee
where the radial potential $V(r)$ could be either
\be
V_{\rm osc}(r)\=\frac{\omega^2r^2}{2}
\qquad\text{or}\qquad
V_{\rm Coul}(r)\=-\frac{\gamma}{r}\ ,
\label{5}
\ee
and ${\cal I}(I_i)$ is the Hamiltonian of some compact $(N{-}1)$-dimensional integrable system
formulated in action-angle variables $(I_i,\Phi_i)$ with $i=1,\ldots,N{-}1$.

The analogous systems on the $N$-sphere $S^N$ and the (two-sheet) hyperboloid $H^N$
of size $r_0$ are
\begin{align}
S^{N}:&\quad {\cal H}=\frac{p_{\chi}^2}{2 r_0^2}+\frac{{\cal I}}{ r_0^2 \sin^2\chi}+V(\tan\chi)\ ,
\qquad
V_{\rm Higgs}=\frac{r^2_0\omega^2}{2}\tan^2\chi\ ,
\qquad
V_{\rm Schr}=-\frac{\gamma}{r_0}\cot\chi\ ,
\label{3}
\\
H^{N}:&\quad {\cal H}=\frac{p_{\chi}^2}{2 r_0^2}+\frac{{\cal I}}{ r_0^2\sinh^2\chi}+V(\tanh\chi)\ ,
\qquad
V_{\rm Higgs}=\frac{r^2_0\omega^2}{2}\tanh^2\chi\ ,
\qquad
V_{\rm Schr}=-\frac{\gamma}{r_0}\coth\chi\ ,
\label{4}
\end{align}
with  $ \{p_\chi, \chi\}=1$ and ${\cal I}(I_i)$ depending on the respective action variables.
The oscillator and Coulomb potentials $V_{\rm higgs}$ and $V_{\rm schr}$
on spheres and hyperboloids have been proposed by Higgs~\cite{higgs} and by Schr\"odinger~\cite{schr}),
respectively.
In other words, we obtain integrable deformations of the $N$-dimensional oscillator and Coulomb systems
by replacing the quadratic $SO(N)$ Casimir invariant with
the Hamiltonian ${\cal I}$ of some $(N{-}1)$-dimensional compact integrable system.

It was shown~\cite{hlnsy} that  the Hamiltonians of the oscillator-like systems are given by
\be
{\cal H}_{\rm osc}\={\cal H}_{\rm osc}(2 I_r + \sqrt{{2\cal I}})
\=
\begin{cases}
\omega (2 I_r + \sqrt{2{\cal I}}) & {\rm for } \quad  \R^{N}\ ,
\\
\frac{1}{2}(2I_\chi+ \sqrt{2{\cal I}}+\omega)^2- \frac{\omega^2}{2} & {\rm for}  \quad S^N\ ,
\\
-\frac{1}{2}(2I_\chi +\sqrt{2{\cal I}}-\omega)^2+ \frac{\omega^2}{2} & {\rm for}  \quad H^N\ .
\end{cases}
\ee
Similarly, the Hamiltonians of the Coulomb-like systems read
\be
{\cal H}_{\rm Coul}={\cal H}_{\rm Coul}( I_r + \sqrt{2{\cal I}})
=
\begin{cases}
-\frac12\gamma^2(I_r + \sqrt{2{\cal I}})^2 & {\rm for }\quad \R^N,
\\
-\frac12\gamma^2 (I_\chi + \sqrt{2{\cal I}})^2+ \frac12 (I_\chi+\sqrt{2{\cal I}})^2 & {\rm for}\quad S^N,
\\
-\frac12\gamma ^2 (I_\chi+\sqrt{2 {\cal I}})^2 - \frac12(I_\chi+\sqrt{2 {\cal I}})^2 & {\rm for}\quad  H^N.
\end{cases}
 \ee
Thus, if the angular Hamiltonian has the form
\be
{\cal I}\=\frac12\Bigl(\sum_i k_i I_i + \text{const} \Bigr)^2 \qquad\text{with}\quad k_i\in \Z\ ,
\label{app}
\ee
then the respective deformations of the oscillator and Coulomb systems are maximally superintegrable.
The deformed systems will have the same configuration space as initial one if ${\cal I}$
is a system on the $(N{-}1)$-sphere.

The key observation in this Note is that any compact system with Hamiltonian (\ref{app}) is precisely
the angular part of a rational Calogero model, which have been suggested and studied in~\cite{sphCal,sph-mat,flp}.
\\

\noindent
{\bf 3. Calogero-Coulomb and Calogero-oscillator models.}
The quantum Hamiltonian of the generalized rational Calogero model associated with a Coxeter root system
\be
\bigl\{\alpha\bigr\}\=\bigl\{(\alpha_1, \alpha_2, \ldots, \alpha_N)\bigr\}\= {\mathcal R}\ \subset\
\R^N \ \ni\ (x_1,x_2,\ldots,x_N) \equiv x
\ee
reads \cite{calogero-root}
\begin{equation}
\label{qHCox}
{\widehat{\cal H}}_{\rm Cal} \=  \sum_i \frac{\hat p_i^2}{2}  +
\sum_{\alpha\in {\mathcal R}_+} \frac{g_\alpha(g_\alpha{-}\hbar)(\alpha\cdot \alpha)}{2(\alpha \cdot x)^2}
\qquad\text{where}\qquad
[{\hat p}_i, x_j]=-{\rm i}\hbar \delta_{ij}\ ,
\end{equation}
${\mathcal R}_+$ is the subset of positive roots, and $g_\alpha\ge 0$ is a multiplicity function
on $\mathcal R$ invariant under the Weyl reflection group~$W$~\cite{Humphreys}.
We indicate quantum objects by `hatting' them.
The Hamiltonian $\widehat{\cal I}$ of the ``angular part'' of ${\widehat{\cal H}}_{\rm cal}$
is defined by rewriting the latter in spherical variables,
\be
{\widehat{\cal H}}_{\rm Cal}\= \frac{{\hat p}_r^2}{2}+\frac{\widehat{\cal I}}{r^2}+
\frac{\hbar^2(N{-}1)(N{-}3)}{8\,r^2}
\qquad\text{with}\quad
r^2= \sum_i x_i^2 \quad\text{and}\quad {\hat p}_r=-{\rm i}\hbar\partial_r \ .
\ee
The basic classical properties of $\widehat{\cal I}$ were considered in~\cite{sphCal,sph-mat,hlnsy},
and its quantum features were investigated in~\cite{flp}.
There, the energy spectrum was found to be
\begin{equation}\label{veCox}
{\cal  E}_{k_2,k_3,k_4,\ldots} \= \sfrac12 q\big(q+\hbar(N{-}2)\big)
\qquad {\rm with} \qquad
q\= \sum_{\alpha \in {\mathcal R}_+} g_\alpha + \hbar  \sum_{i=2}^N d_i k_i,
\qquad {\rm and}\qquad k_i=0,1,2,\ldots\  .
\end{equation}
Here, $d_1{=}2, d_2, \ldots, d_N$ are the degrees of the basic homogeneous $W$-invariant
polynomials $\sigma_1{=}r^2, \sigma_2, \ldots, \sigma_N$.
Note that the quantum number $k_1$ from the contribution of $\sigma_1$ is absent in the energy formula.

Since in the classical limit $\hbar\to 0$ the quantum numbers $\hbar k_i$ reduce to the action variables $I_i$,
the above relation yields the classical Hamiltonian in terms of action variables:
\begin{equation}
\label{ECox}
{\cal I} \= \lim_{\hbar\to0}\sfrac12 q^2\=
\frac12 \biggl( \sum_{i=2}^N d_i I_i + \sum_{\alpha \in {\mathcal R}_+} g_\alpha \biggr)^2\ .
\end{equation}
In particular, the angular part of rational $A_{N-1}$ Calogero model is given by
\be
 {\cal I}\=\sfrac12 \bigl( \sfrac{N(N{-}1)}{2} g +I_1+3I_3+\ldots+N I_N \bigr)^2\ ,
\label{sphCal}
\ee
Note that, compared to \eqref{ECox}, the index $i$ has been shifted up by one, and there is
an additional contribution of $I_1$ from the center of mass
which adds an extra $A_1$ to the $A_{N-1}$ root system.
We see that the angular part of the generalized rational Calogero model belongs to the class \eqref{app},
i.e.~it is suitable for the construction of superintegrable deformations of oscillator and Coulomb systems.

Substituting \eqref{sphCal} into \eqref{2}-\eqref{4},
we obtain classical maximally superintegrable deformations of (Euclidean, spherical and pseudospherical)
oscillator and Coulomb systems.
In Euclidean space $\R^N$, these systems are then characterized by the Hamiltonians
\be
\label{Hcou}
{\cal H}_{\rm osc} \= \frac{p^2}{2}  + \sum_{\alpha\in {\mathcal R}_+}
\frac{g^2_\alpha(\alpha\cdot \alpha)}{2(\alpha\cdot x)^2} +\frac{\omega^2 x^2}{2}
\und
{\cal H}_{\rm Coul} \= \frac{p^2}{2} + \sum_{\alpha\in {\mathcal R}_+}
\frac{g^2_\alpha(\alpha\cdot \alpha )}{2(\alpha\cdot x)^2} - \frac{\gamma}{x}\ ,
\ee
where $p\equiv(p_1,p_2,\ldots,p_N)$ and $x\equiv(x_1,x_2,\ldots,x_N)$.
The maximal superintegrability is well-known for the oscillator case, due to the knowledge of
oscillating quantities with integer frequencies in the harmonic potential~\cite{calogero-root,gonera99-2}.
The Calogero-Coulomb system seems much less known~\cite{khare,comment},\footnote{
In these papers only the rational $A_{N-1}$ Calogero model was considered.
However, the generalization to arbitrary Coxeter root systems is straightforward.}
and its superintegrability has not been discussed in the literature yet.
Here we have established that the generalized rational Calogero-Coulomb model for any Coxeter root system
is maximally superintegrable.

The story extends from Euclidean space to spaces of constant curvature,
i.e.~to the $N$-dimensional sphere and pseudosphere,
which we present in terms of Euclidean coordinates on the ambient space $\R^{N+1}\ni(x_0,x)$,
\be
\label{qHS}
\begin{gathered}
{\cal H}^{(p)s}_{\rm Coul}\=\frac{p^2}{2} \mp \frac{(x\cdot p)^2}{2r^2_0}+
\sum_{\alpha \in {\mathcal R}_+}
\frac{g_\alpha^2(\alpha\cdot \alpha )}{2(\alpha\cdot x)^2}\
-\ \frac{\gamma}{r_0}\frac{x_0}{x}
\quad\qquad\text{with}\qquad
x^2_0 \pm x^2= r^2_0\ ,
\\
{\cal H}^{(p)s}_{\rm osc}\=\frac{p^2}{2} \mp \frac{(x\cdot p)^2}{2r^2_0}+
\sum_{\alpha \in {\mathcal R}_+}
\frac{g_\alpha^2(\alpha\cdot \alpha ) }{2(\alpha\cdot x)^2}\
+\ \frac{\omega^2r^2_0}{2}\frac{x^2}{x_0^2}
\qquad\text{with}\qquad
x^2_0 \pm x^2= r^2_0\ .
\end{gathered}
\ee
In these expressions, the upper sign corresponds to the sphere and lower sign  to the hyperboloid.

As in the Euclidean case, these systems are superintegrable.
Their potentials may be viewed as a higher-dimensional  generalizations of superintegrable deformations
of two-dimensional oscillator and Coulomb potentials which are known as
Tremblay-Turbiner-Winternitz~\cite{TTW} and Post-Winternitz~\cite{PW} systems, respectively.
Generalizations of the latter to spheres and pseudospheres have been proposed in~\cite{hlnsy}.
We note that the two-dimensional Post-Winternitz system possesses just three constants of motion,
which are all quadratic in the momenta. This fact implies that it admits a separation of variables
in some coordinate system different from the spherical one.
Since the Tremblay-Turbiner-Winternitz and Post-Winternitz systems are connected
by a Bohlin-Levi-Civita transformation, the same separability statement holds
for the Tremblay-Turbiner-Winternitz system as well.
\\

\noindent
{\bf 4. Quantum mechanics.}
Let us briefly consider the quantum mechanics of the proposed systems.
The quantum versions of the Hamiltonians (\ref{qHCox}) and (\ref{qHS}) are obtained by replacing
their kinetic term by (minus) the Laplacian and by deforming the couplings
$g^2_\alpha$ to $g_\alpha(g_\alpha{-}\hbar)$.
If the $g_\alpha$ are integer multiples of $\hbar$, then the system is even algebraically integrable~\cite{clp}.

It is then easy to deduce that the wavefunctions of the Calogero-Oscillator and Calogero-Coulomb systems
are given by
\be
\Psi_{n_rq}(x_i)\=\psi_{n_rq}(x_i/r)R_{n_rq}(r)\ ,
\ee
where $R_{n_rq}(r)$ are the radial wavefunctions of the corresponding oscillator and Coulomb problems
with orbital quantum number $\ell$ having been replaced with $q$ as given by~\eqref{veCox}.
For the systems on spheres and hyperboloids the role of the radius $r$ is played by the angle
$\arccos(x_{N+1}/r_0)$. The expressions for these radial wavefuctions are well known~\cite{pogosyan0}.
Analogously, the energy spectra of these systems are found by replacing $\ell\to q$ in the spectra of
the oscillator and Coulomb systems (here we put $\hbar{=}1$),
\be
\begin{aligned}
E_{\rm osc} &\=\Bigl(\sqrt{\omega^2+\sfrac14r^{-4}_0}-\sfrac12r^{-2}_0\Bigr)\bigl(2n_r+q+\sfrac{N}{2}\Bigr)
\pm\sfrac12r^{-2}_0\bigl(2n_r+q+\sfrac12\pm\sfrac12\bigr)\bigl(2n_r+q+N+\sfrac12\mp\sfrac12\bigr)\ ,
\\[4pt]
E_{\rm Coul} &\=-\sfrac12\gamma^2\bigl(n_r+q+\sfrac{N{-}1}{2}\bigr)^{-2}
\pm\sfrac12r^{-2}_0(n_r+q)(n_r+q+n-1)\ ,
\end{aligned}
\ee
where, as before, the upper sign corresponds to the sphere and lower one to the two-sheet hyperboloid.
In the latter case, requiring positivity restricts the principal quantum numbers.
In the situation where $q$ is an integer multiple of~$\hbar$, the two spectra agree with the undeformed ones,
i.e.~with the $SO(N)$ quadratic Casimir instead of the angular Calogero Hamiltonian,
but the degeneracies are different.
For the Calogero-oscillator system, the energy level is determined by the principal number $n=2n_r+q$,
while for the Calogero-Coulomb system it is given by $n=n_r+q$, where $q$ is defined in~(\ref{veCox}).
Thus, the  degeneracy of the $n$-th level depends on the corresponding Coxeter root system.
\\

\noindent
{\bf 5. Matrix model reduction.}
The superintegrability of the Calogero-Coulomb model can easily be understood in terms of a matrix model.
For simplicity, we consider here the root system of $u(N)\simeq u(1)\oplus su(N)$, which leads to the
standard rational Calogero model \cite{calogero0}, although the results can be extended to any root system.

The Hermitean matrix model for a particle in a Coulomb potential is defined by the Hamiltonian
\be
\label{Hmat}
{\cal H}_\text{mat} \= \sfrac12 \text{tr}\,\mathbf{P}^2 - \gamma\left(\text{tr}\,\mathbf{X}^2\right)^{-\frac12}
\=\sfrac12 \smash{\sum_a} P_a^2 - \frac{\gamma}{r}
\qquad\text{with}\quad r^2 = \smash{\sum_a} X_a^2 \und a=0,1,\ldots N^2{-}1\ .
\ee
Here $\mathbf{P}$ and $\mathbf{X}$ denote
Hermitean matrices containing $N^2$ momenta~$P_a$ and coordinates~$X_a$, respectively,
\be
\mathbf{P} = \smash{\sum_a} P_a T_a \und \mathbf{X} = \smash{\sum_a} X_a T_a\ ,
\ee
where we introduced a basis of $U(N)$ generators $T_a$ orthonormalized as $\text{tr}\, T_aT_b=\delta_{ab}$.

It is well known that the integrals of motion of such a system are given
by the angular momentum tensor and the Runge-Lenz vector,
\begin{gather}
\label{matL}
\mathbf{M} \= \mathbf{X}\otimes\mathbf{P}-\mathbf{P}\otimes\mathbf{X}
\= \sum_a M_{ab}\ T_a\otimes T_b \qquad\text{with}\quad M_{ab}=X_aP_b-X_bP_a\ ,
\\
\label{matA}
\mathbf{A} \= (\unity\otimes\text{tr})\mathbf{M}(\unity\otimes\mathbf{P})-\frac{\gamma}{r}\mathbf{X}
\= \Bigl(\text{tr}\,\mathbf{P}^2-\frac{\gamma}{r}\Bigr)\mathbf{X} - \text{tr}(\mathbf{X}\mathbf{P})\,\mathbf{P}
\= \smash{\sum_a} A_a T_a \ ,
\end{gather}
so that the components of the Runge-Lenz vector acquire the standard form
\be
\label{Aa}
A_a \= \smash{\sum_b} M_{ab}P_b - \frac{\gamma}{r} X_a
\= \Bigl(\smash{\sum_b} P_b^2-\frac{\gamma}{r}\Bigr) X_a - \Bigl(\smash{\sum_b} X_b P_b\Bigr) P_a\ .
\ee
The matrix Hamiltonian \eqref{Hmat} is preserved under the adjoint $SU(N)$ action.
For $\gamma=0$, the corresponding reduction gives rise to the Calogero Hamiltonian
\cite{kazhdan,calogero-root,calogero-review}.
The coordinate and momentum matrices are reduced respectively to
\be
\label{XPred}
\mathbf{X}\ \to\ \bigl(X_{ij}\bigr)\=\bigl(x_i\delta_{ij}\bigr) \und
\mathbf{P}\ \to\ \bigl(P_{ij}\bigr)\=\Bigl(p_i\delta_{ij}+\im g\frac{1-\delta_{ij}}{x_i-x_j}\Bigr)\ .
\ee
For nonvanishing $\gamma$, this reduction procedure leads to the Hamiltonian ${\cal H}_\text{Coul}$
of the Calogero model with a Coulomb potential \eqref{Hcou},
since the potential term in the matrix Hamiltonian \eqref{Hmat} is also invariant
with respect to the adjoint action.

From the integrals of motion \eqref{matL} and \eqref{matA} of the matrix model,
only the $SU(N)$-invariant ones survive the reduction procedure.
Such invariants can be constructed using $SU(N)$-invariant tensors.
In \cite{sph-mat} the general form of such invariants, containing angular momentum tensors only,
is described using a graphical representation. They correspond to the invariants of the
angular Hamiltonian $\mathcal{I}$ defined in \eqref{2}. Among them are the traces of even powers
of the angular momentum matrix,
\be
\label{Ln}
\mathcal{M}_n\=(\text{tr}\otimes\text{tr})\,\mathbf{M}^{2n}\ .
\ee

The aforementioned graphical picture may be extended by including the Runge-Lenz vector \eqref{matA}.
In particular, the quantities
\be
\label{An}
\mathcal{A}_n\=\text{tr}\,\mathbf{A}^n
\ee
are $SU(N)$ invariants and, hence, yield integrals of motion in the reduced model.
The matrix \eqref{matA} reduces to the explicit form
\be
\mathbf{A}\ \to\ \Bigl(2{\cal H}_\text{Coul}+\frac{\gamma}{r}\Bigr) \mathbf{X} - rp_r\mathbf{P}\ ,
\ee
where $\mathbf{X}$ and $\mathbf{P}$ are now defined by \eqref{XPred}, and
\be
{\cal H}_\text{Coul} \= \smash{\sum_i} \frac{p_i^2}{2} + 
\smash{\sum_{i<j}}\frac{g^2}{(x_i{-}x_j)^2}-\frac{\gamma}{r}\ .
\ee
The first integral from the series \eqref{An} is
\be
\label{A1}
\mathcal{A}_1 \= \Bigl(2{\cal H}_\text{Coul}+\frac{\gamma}{r}\Bigr)  \sum_i x_i  - rp_r\sum_i p_i \ .
\ee

The quantum system can be treated using the Dunkl operators \cite{dunkl}.
We employ their gauged $SU(N)$ version,
\be
\label{dunkl}
\nabla_i\=\partial_i- \smash{\sum_{k(\ne i)}} \frac{g}{x_i{-}x_k}s_{ik}\ ,
\ee
where $s_{ik}$ permutes the $i$th and $k$th particle labels.
Like ordinary derivatives, the operators $\nabla_i$ mutually commute
and obey nontrivial commutation relations with the coordinates $x_i$,
\be
\label{com-nabla}
[\nabla_i,x_j]\=
\begin{cases}
 - g s_{ij} & \text{for $i\ne j$}\ ,
\\[4pt]
 1+g\sum_{k(\ne i)}s_{ik} & \text{for $i=j$}\ .
\end{cases}
\ee

The Dunkl operators serve as a tool for the construction and study of the Calogero model
and its various extensions \cite{brink-poly,calogero-review}.
In particular, is is easy to see that the restriction of the Hamiltonian
\be
\label{Hcal-dunkl}
-\sfrac12\mathbf{\nabla}^2\big|_{\text{sym}} \=\widehat{\mathcal{H}}_\text{Cal}
\ee
on totally symmetric wavefunctions produces the Calogero model.
In order to construct the integrals of motion for
\be
\label{Hcou-dunkl}
\widehat{\mathcal{H}}_\text{Coul} \= \widehat{\mathcal{H}}_\text{Cal} - \frac{\gamma}{r}\ ,
\ee
we consider the Dunkl {\sl angular\/} momentum operators investigated in detail in \cite{fh}:
\be
\label{Lij}
\widehat M_{ij}\= x_i\nabla_j-x_j\nabla_i\ .
\ee
Such operators commute with $\widehat{\mathcal{H}}_\text{Cal}$
and its angular part $\widehat{\cal I}$ \cite{fh}.
Therefore, they commute with $\widehat{\mathcal{H}}_\text{Coul}$ \eqref{Hcou-dunkl} too,
since they depend on the angular coordinates only.
Because we restrict to totally symmetric wavefunctions, we have to symmetrize \eqref{Lij}
in order to get the true integrals:
\be
\label{qLn}
\widehat{\mathcal{M}}_n\=\smash{\sum_{i<j}}\bigl(\widehat{M}_{ij}\bigr)^{2n}\big|_\text{sym}
\ee
are the quantum analogs of the integrals \eqref{Ln}.
Note that $\widehat{\mathcal{M}}_2$ is proportional to $\widehat{\cal I}$ up to a constant.

In complete analogy with \eqref{Lij},
one can define the quantum Runge-Lenz operator expressed via Dunkl ones as
\be
\widehat A_i \=-\sfrac12\smash{\sum_{k}}\{\widehat M_{ik},\nabla_k\} - \frac{\gamma x_i}{r}\ .
\ee
The anticommutator in this expression guarantees its hermiticity.
However, such operators do not commute with the Hamiltonian \eqref{Hcou-dunkl}.
Nevertheless, their sum
\be
\widehat{\cal A}_1 
\=-\sfrac12\smash{\sum_{i,k}}\{\widehat M_{ik},\nabla_k\}-\frac{\gamma}{r}\smash{\sum_i} x_i
\= \Bigl(\smash{\sum_i} x_i\Bigr)\Bigr(2\widehat{\mathcal H}_\text{Coul}+\frac{\gamma}{r}\Bigr) 
+ \bigl(r\partial_r + \sfrac{N-1}{2}\bigr)\smash{\sum_i} \partial_i
\label{RL}
\ee
is a constant of motion of the system, as can be verified by direct calculation \footnote{
For $\gamma{=}0$ it is nothing but the first Wojciechowski integral named $F_1$ in \cite{clp}.}.
The final expression in \eqref{RL} is a quantum version of the classical expression \eqref{A1}.
The quantities $\widehat{\mathcal{H}}_\text{Coul}$ and $\widehat{\cal A}_1$ extend the
$2N{-}3$ conserved quantities of the angular subsystem $\widehat{\cal I}$
to the full number $2N{-}1$ of conserved charges required for superintegrability.
\\

\noindent
{\bf 6. Concluding remarks.}
In this note we announce the maximal superintegrability of the generalized rational Calogero model
(based on any Coxeter root system) perturbed by a Coulomb potential, on Euclidean $\R^N$ as well as
on a sphere~$S^N$ or pseudosphere~$H^N$. The same feature holds true for a harmonic oscillator potential
in place of the Coulomb one.
In fact, we demonstrated that the Calogero-Coulomb and Calogero-oscillator systems at integer values
of the coupling are the \emph{only} isospectral deformations of the standard oscillator and Coulomb systems.
It should be noted that, except for the Euclidean $A_{N-1}$ Calogero-oscillator system, these models cannot be
naturally interpreted as $N$-particle systems because of nonstandard multi-particle interactions.
\\

\noindent
{\bf Acknowledgments.}
The authors thank A.~Polychronakos for encouragement at the initial stage of this investigation.
This work was partially supported by the Volkswagen Foundation grant I/84 496,
by the Armenian State Committee of Science grant SCS 13-1C114,
and by ANSEF grant 3501-math-phys (T.H., A.N.).

\end{document}